\documentclass[runningheads,12pt]{llncs}
\usepackage{float}
\usepackage{booktabs}
\usepackage{graphicx}
\usepackage{amsmath,amssymb}
\usepackage{hyperref}
\usepackage{multirow}
\usepackage{xcolor}
\usepackage[margin=2.5cm]{geometry}
\usepackage{setspace}
\usepackage{subcaption}
\title{A Multi-Modal Spatial Risk Framework for EV Charging Infrastructure Using Remote Sensing}

\author{Oktay Karaku\c{s}\orcidID{0000-0001-8009-9319} \and
Padraig Corcoran\orcidID{0000-0001-9731-3385}}
\authorrunning{O. Karaku\c{s} \& P. Corcoran}
\institute{Cardiff University, School of Computer Science and Informatics, UK\\
\email{\{karakuso,corcoranp\}@cardiff.ac.uk}}
\begin{document}
\maketitle

\begin{abstract}
Electric vehicle (EV) charging infrastructure is increasingly critical to sustainable transport systems, yet its resilience under environmental and infrastructural stress remains underexplored. In this paper, we introduce RSERI-EV, a spatially explicit and multi-modal risk assessment framework that combines remote sensing data, open infrastructure datasets, and spatial graph analytics to evaluate the vulnerability of EV charging stations. RSERI-EV integrates diverse data layers, including flood risk maps, land surface temperature (LST) extremes, vegetation indices (NDVI), land use/land cover (LULC), proximity to electrical substations, and road accessibility to generate a composite Resilience Score. We apply this framework to the country of Wales EV charger dataset to demonstrate its feasibility. A spatial $k$-nearest neighbours ($k$NN) graph is constructed over the charging network to enable neighbourhood-based comparisons and graph-aware diagnostics. Our prototype highlights the value of multi-source data fusion and interpretable spatial reasoning in supporting climate-resilient, infrastructure-aware EV deployment.
\end{abstract}

\section{Introduction}

The rapid expansion of electric vehicle (EV) infrastructure is central to national decarbonisation efforts and the transition to sustainable transport. Public and private actors across Europe (and globally) are investing heavily in deploying EV charging stations, aiming to improve accessibility and encourage EV adoption \cite{lamonaca2022state}. However, much of this infrastructure remains exposed to a growing range of environmental and infrastructural risks. Flooding, extreme heat events, network congestion, and poor road accessibility can all compromise the operational continuity and equity of EV networks \cite{torkey2024transportation,raman2022resilience}. As the climate crisis intensifies, ensuring the spatial resilience of EV charging infrastructure is a critical challenge for local authorities, grid operators, and urban planners.

While existing work in this domain has largely focused on optimising charger placement based on demand or grid load balancing \cite{ccelik2024electric,yuvaraj2024comprehensive}, relatively little attention has been paid to the environmental vulnerability of deployed infrastructure \cite{mohammed2024strategies,sayarshad2024optimization}. Few frameworks integrate multi-modal environmental and infrastructural stressors into a unified decision-support tool. At the same time, advances in open remote sensing (RS) and geospatial analytics offer new opportunities for scalable, data-driven vulnerability assessment. Products such as Sentinel-2 and Landsat 8 provide access to detailed, regularly updated imagery on land surface temperature (LST), vegetation stress, and land use \cite{halder2025remote}. Combined with vector datasets on flood zones, road access, and power grid topology, these can form the backbone of spatial risk modelling for EV infrastructure.

In this paper, we introduce \textbf{RSERI-EV}, a prototype framework for assessing spatial risk to EV charging stations through the fusion of environmental hazard data, infrastructure proximity metrics, and RS layers. Our approach incorporates multiple open datasets and RS products to derive a composite \textit{Resilience Score} for each EV charger, using features such as flood zone overlap, LST anomaly, substation distance, vegetation index, and land use classification. The framework also includes a spatial $k$-nearest neighbour ($k$NN) graph constructed over charger locations to support optional neighbourhood-level comparisons and spatial aggregation. While the current use of graph topology is exploratory and intentionally light-weight, it demonstrates the potential for spatial reasoning and diffusion-based metrics for resilience forecasting.

We apply this framework to a regional EV charger dataset, focusing on the country of Wales as a case study. Our pipeline fuses remote sensing and infrastructure data into spatial risk layers and computes a resilience score (RSERI) for each charger. While this initial work emphasises integration and descriptive analysis, the framework is designed to accommodate further extensions, including predictive modelling, stakeholder feedback, and reasoning via graph structures. To our knowledge, this is among the first end-to-end efforts to combine EO-driven hazards and infrastructure vulnerability for EV charging networks.

The rest of the paper is structured as follows: Section~\ref{sec:related} reviews literature across EV infrastructure risk modelling and geospatial data fusion. Section~\ref{sec:method} details the data sources, feature engineering process, and spatial graph construction. Section~\ref{sec:results} presents results, visualisations, and insights. Section~\ref{sec:discussion} discusses key implications, limitations, and future extensions. We conclude in Section~\ref{sec:conclusion}.

\section{Related Works}\label{sec:related}
The resilience of EV charging infrastructure has garnered increasing attention, particularly in the context of environmental and infrastructural stressors. Traditional approaches to EV charging station (EVCS) placement have primarily focused on optimising for demand, cost, and accessibility. However, recent studies have begun to incorporate environmental risks and multi-modal data sources to enhance the robustness of EVCS networks.


Recent research emphasises the integration of environmental factors into EVCS planning. For instance, Batic et al.~\cite{batic2025geodemographic} introduced a geodemographic-aware model that considers socio-economic disparities in charger placement, highlighting the need for equitable infrastructure development. Similarly, Almutairi~\cite{almutairi2022reliability} assessed the impact of diverse EVCS on overall service reliability, underscoring the importance of considering various charging scenarios and their implications on power systems. RS technologies have also been leveraged to assess environmental risks. Prakash et al.~\cite{charly2024evremote} utilised multi-sensor RS data to identify urban locations for EVCS deployment, demonstrating the potential of RS in informing infrastructure planning. Moreover, Harb and Dell’Acqua~\cite{harb2017remotesensing} discussed the role of RS in multi-risk assessment, providing insights into how RS can contribute to disaster management and infrastructure resilience.


Banegas and Mamkhezri~\cite{banegas2023gisreview} conducted review of GIS-based methods, revealing that 
the Analytic Hierarchy Process (AHP) and Technique for Order Preference by Similarity to Ideal Solution (TOPSIS) are prevalent in evaluating potential EVCS locations. These approaches often consider factors such as proximity to roads, population density, and environmental constraints. However, the integration of environmental risk factors into Multi-criteria decision-making (MCDM) frameworks remains limited. The RSERI-EV addresses this gap by incorporating diverse environmental and infrastructural data layers, enabling a comprehensive assessment of EVCS resilience.


Graph-based methods have emerged as powerful tools for modelling spatial relationships in EV infrastructure. Batic et al.~\cite{batic2025geodemographic} employed a Graph Neural Network (GNN) to capture the interconnected nature of urban zones, facilitating more informed charger placement decisions. Similarly, Zhang et al.~\cite{zhang2024mstem} proposed a Multiscale Spatio-Temporal Enhanced Model (MSTEM) for short-term load forecasting at EVCS, utilising a multiscale GNN to capture spatiotemporal variations in charging patterns. While these studies demonstrate the potential of graph-based approaches, their application in environmental risk assessment for EVCS is still evolving. The RSERI-EV incorporates a spatial kNN graph to enable neighbourhood-level comparisons and spatial aggregation, offering a novel perspective on EVCS resilience analysis.


The fusion of RS data with graph analytics presents a promising avenue for comprehensive EVCS resilience assessment. Li et al.~\cite{li2022deeprsfusion} reviewed deep learning techniques for multimodal RS data fusion, highlighting the potential for integrating heterogeneous data sources in geospatial applications. In the context of EV infrastructure, such integration can facilitate the identification of high-risk areas and inform strategic planning. The RSERI-EV framework exemplifies this integration by combining RS-derived environmental indicators with graph-based spatial analysis, enabling a multi-faceted evaluation of EVCS vulnerability. This approach aligns with the growing emphasis on data-driven decision-making in infrastructure planning and climate resilience.

Overall, while prior work has made significant strides in site selection, environmental analysis, and graph-based spatial modelling, few studies have holistically combined these dimensions within a unified, data-driven framework. This paper addresses that gap by proposing RSERI-EV, a modular and multi-modal resilience scoring system for EV charging infrastructure. By fusing Earth Observation data, proximity-based environmental and infrastructural indicators, and a spatial graph abstraction, we contribute an interpretable yet extensible framework suitable for both research and policy applications. 

\section{Methodology}\label{sec:method}

RSERI-EV is a modular framework designed to assess the spatial resilience of EV charging infrastructure by integrating multi-modal environmental and infrastructural data. The framework computes a composite Resilience Score for each charging station, incorporating diverse risk indicators. Additionally, a spatial graph is constructed to facilitate neighbourhood-level analyses and potential future applications such as counterfactual reasoning and spatial smoothing.

\subsection{Data Sources and Preprocessing}

\textbf{EV Charging Station Data:} Publicly available data from OpenChargeMap was utilised, encompassing charger locations, types, and operational statuses. Data preprocessing involved filtering for active chargers, and no additional geocoding service was used.

\vspace{0.1cm}\noindent\textbf{Environmental Risk Layers:}
\begin{itemize}
    \item \textit{Flood risk }shapefiles were sourced from the Environment Platform Wales (EPW), identifying areas susceptible to flooding.
    \item \textit{Land Surface Temperature (LST)} data were acquired from Landsat 8 via Google Earth Engine (GEE) from April to October 2024, focusing on maximum and median temperatures to detect anomalies.
    \item \textit{Normalised Difference Vegetation Index (NDVI)} values were derived from Sentinel-2 imagery for the same LST time period, classifying vegetation health into low, moderate, and high categories.
\end{itemize}

\vspace{0.1cm}\noindent\textbf{Infrastructural Features:}
\begin{itemize}
    \item \textit{Substation Proximity:} Locations of electrical substations were obtained from OpenStreetMap (OSM). Euclidean distances from each charger to the nearest substation were calculated using the British National Grid projection (EPSG:27700), which offers locally accurate measurements for Wales. While more precise measures such as haversine or network distances could be used, Euclidean distance in projected space provides a robust proxy given the regional scale and uniformity of the CRS.
    \item \textit{Road Network Accessibility:} Major road networks, including motorways and primary roads, were extracted from OSM. Proximity analyses determined the accessibility of each charging station.
\end{itemize}

\subsection{Spatial Risk Indicators}

Each charging station was evaluated against several risk indicators:

\begin{itemize}
    \item \textbf{Flood Risk:} Binary indicator denoting whether a charger is located within a designated flood zone.
    \item \textbf{LST Risk:} Chargers exhibiting land surface temperatures exceeding the regional 90th percentile were flagged as high-risk.
    \item \textbf{Grid Risk:} Chargers located more than 5 km from the nearest substation were identified as having higher grid vulnerability.
    \item \textbf{Road Risk:} Chargers with limited accessibility to major roads were marked as higher risk.
    \item \textbf{NDVI Risk:} Chargers situated in areas with low NDVI values were considered at higher risk due to poor vegetation health.
    \item \textbf{Land Use/Land Cover (LULC) Risk:} Chargers in urban or coastal areas, as determined by land cover classifications, were considered more susceptible to environmental stressors.
    \item \textbf{Vegetation Risk:} Chargers in Low NDVI regions and Urban or Vegetation LULC regions were set as higher risk.
\end{itemize}

\subsection{Graph Construction}

A spatial $k$-nearest neighbors (kNN) graph was constructed to model the spatial relationships among charging stations:

\begin{itemize}
    \item \textbf{Nodes:} Each EVCS represents a node in the graph.
    \item \textbf{Edges:} Edges connect each node to its $k=5$ nearest neighbors based on Euclidean distance.
\end{itemize}

This graph structure facilitates the analysis of local spatial dependencies and can be extended to incorporate advanced spatial analytics in future work.

\subsection{RSERI Score Computation}

The Resilience Score for each charging station was computed as follows:

\begin{itemize}
    \item \textbf{Normalisation:} Continuous risk indicators were normalised to a common scale.
    \item \textbf{Aggregation:} A composite score was calculated by aggregating the normalised risk indicators. Equal weighting was applied in the baseline model, with optional weighting schemes derived from Principal Component Analysis (PCA) explored for sensitivity analysis.
    \item \textbf{Handling Missing Data:} Stations with incomplete data were excluded from the analysis to maintain data integrity.
\end{itemize}

\section{Experimental Analysis \& Results}\label{sec:results}
This section presents the spatial risk analysis results generated using the RSERI-EV framework. We evaluate both individual risk factors and their intersections, explore spatial graphs and regional patterns, and present the distribution of composite RSERI scores across Wales.


Table~\ref{tab:risk_summary} summarises the distribution of EV charging stations (EVCS) across five binary risk layers. LST and vegetation risks are the most prevalent, with 72.5\% and 52.5\% of stations respectively exposed. In contrast, grid and road proximity risks affect a much smaller subset (1.8\% and 7.8\%). Table~\ref{tab:risk_summary2} provides insights into logical intersections among risk layers. Notably, 110 EVCS are exposed to both flood and heat risks, while 376 are impacted by both LST and vegetation stress. Only a small number of chargers (8) fall within both grid and road risk zones, indicating relatively independent exposure types.

\begin{table}[htbp]
\centering
\caption{Distribution of EVCS across individual binary risk factors.}
\label{tab:risk_summary}
\begin{tabular}{lccc}
\toprule
\textbf{Risk Factor} & \textbf{High Risk (1)} & \textbf{Low Risk (0)} & \textbf{High Risk (\%)} \\
\toprule
Flood Risk&141&779&15.3\%\\
LST Risk&667&253&72.5\%\\
Grid Risk&17&903&1.8\%\\
Road Risk&72&848&7.8\%\\
Vegetation Risk&483&437&52.5\%\\
\hline
At least 1 Risk&817&103&88.8\%\\
\bottomrule
\end{tabular}
\centering
\caption{Logical intersections of risk exposure among EVCS.}
\label{tab:risk_summary2}
\begin{tabular}{lccc}
\toprule
\textbf{Risk Factor} & \textbf{High Risk (1)} & \textbf{Low Risk (0)} & \textbf{High Risk (\%)} \\
\toprule
Flood $\cap$ LST&110&810&11.9\%\\
Flood $\cap$ Vegetation&85&835&9.2\%\\
LST $\cap$ Vegetation&376&544&40.9\%\\
Grid $\cap$ Road&8&912&0.9\%\\
Flood $\cap$ LST $\cap$ Vegetation&66&854&7.2\%\\
\bottomrule
\end{tabular}
\end{table}

\begin{figure}[ht!]
    \centering
    \begin{minipage}[b]{0.32\textwidth}
        \centering
        \includegraphics[width=\textwidth]{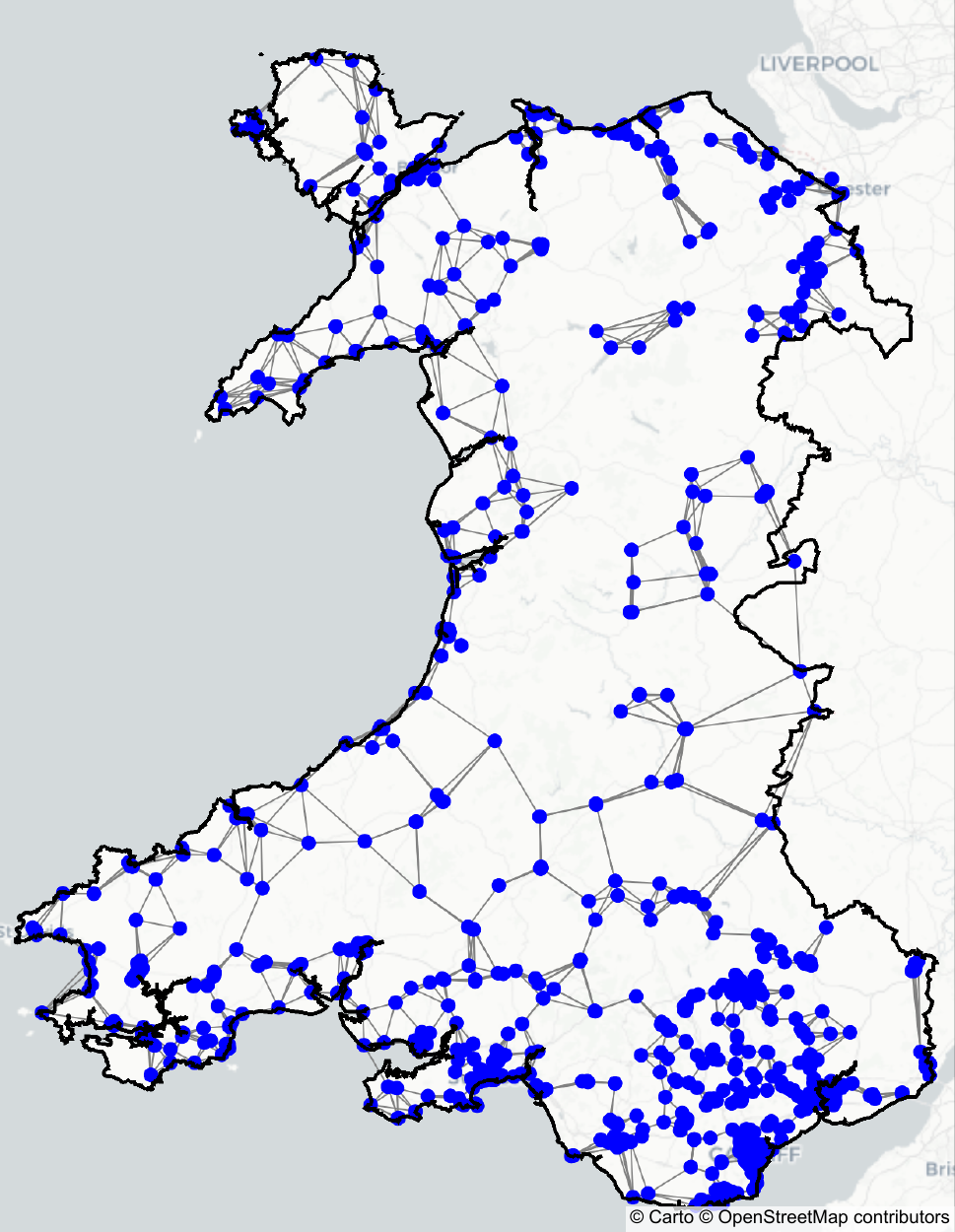}
        \subcaption{EVCS Graph View}
    \end{minipage}
    \hfill
    \begin{minipage}[b]{0.32\textwidth}
        \centering
        \includegraphics[width=\textwidth]{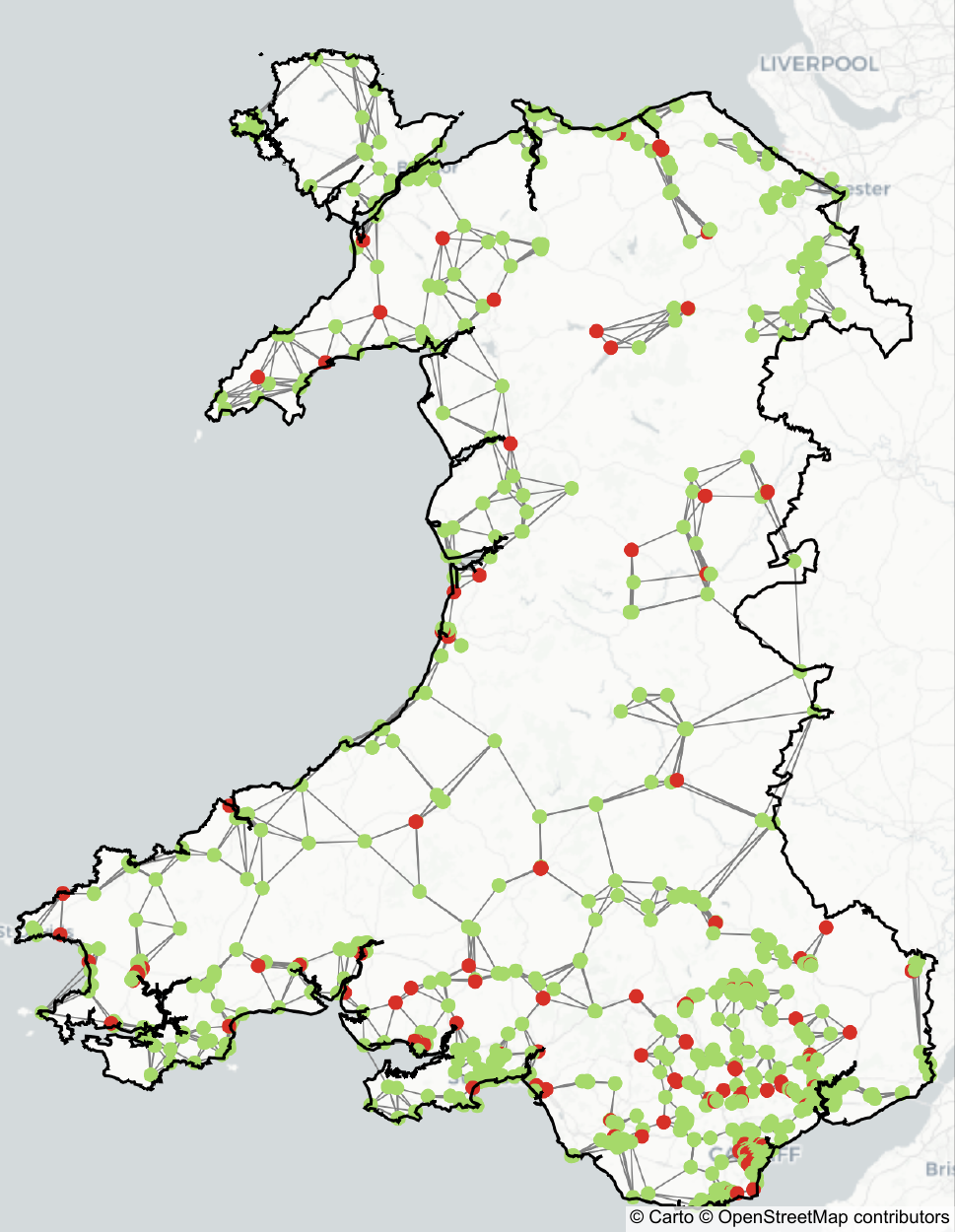}
        \subcaption{Flood}
    \end{minipage}
    \hfill
    \begin{minipage}[b]{0.32\textwidth}
        \centering
        \includegraphics[width=\textwidth]{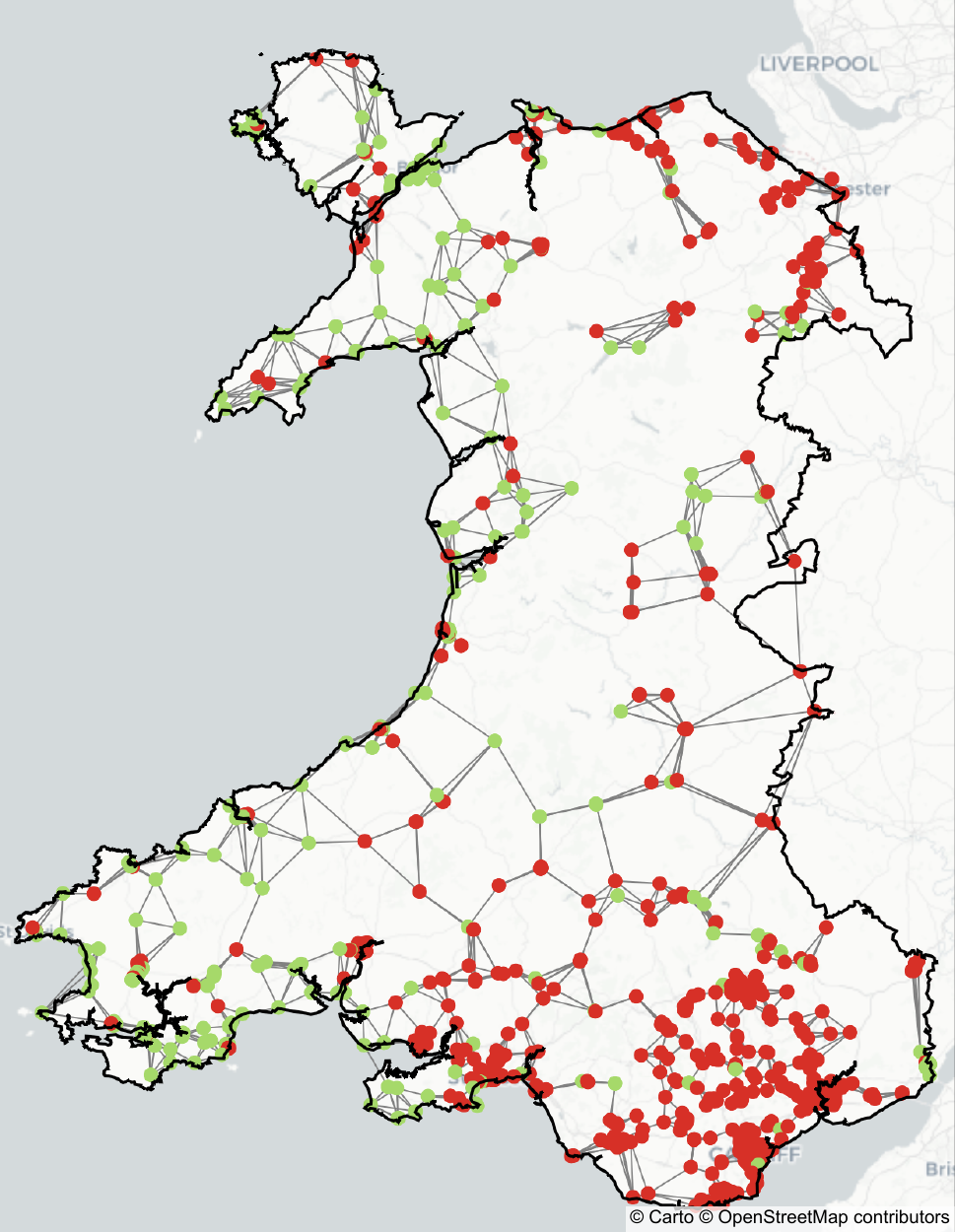}
        \subcaption{LST}
    \end{minipage}
    \vspace{0.4cm}
    
    \begin{minipage}[b]{0.32\textwidth}
        \centering
        \includegraphics[width=\textwidth]{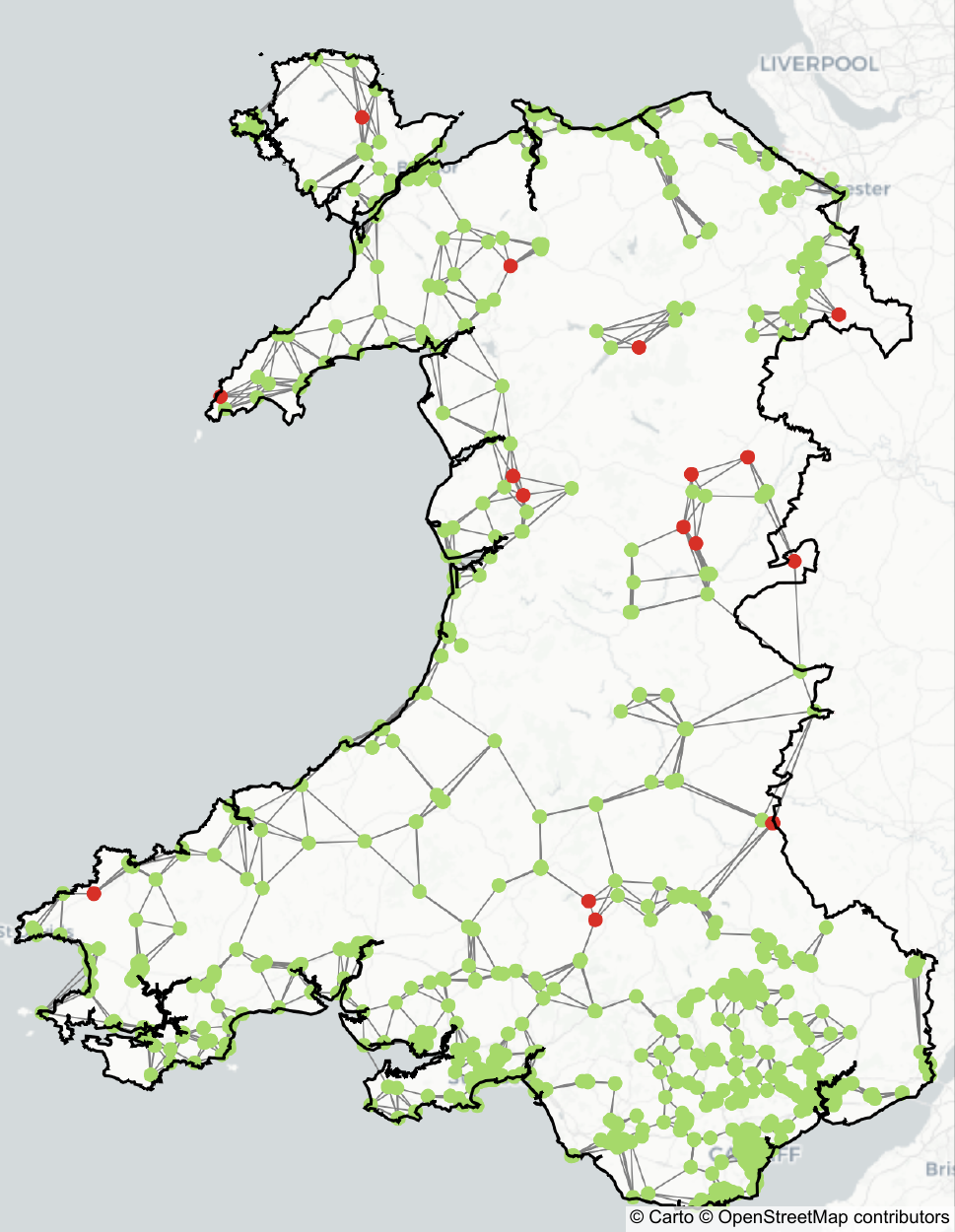}
        \subcaption{Grid Proximity}
    \end{minipage}
    \hfill
    \begin{minipage}[b]{0.32\textwidth}
        \centering
        \includegraphics[width=\textwidth]{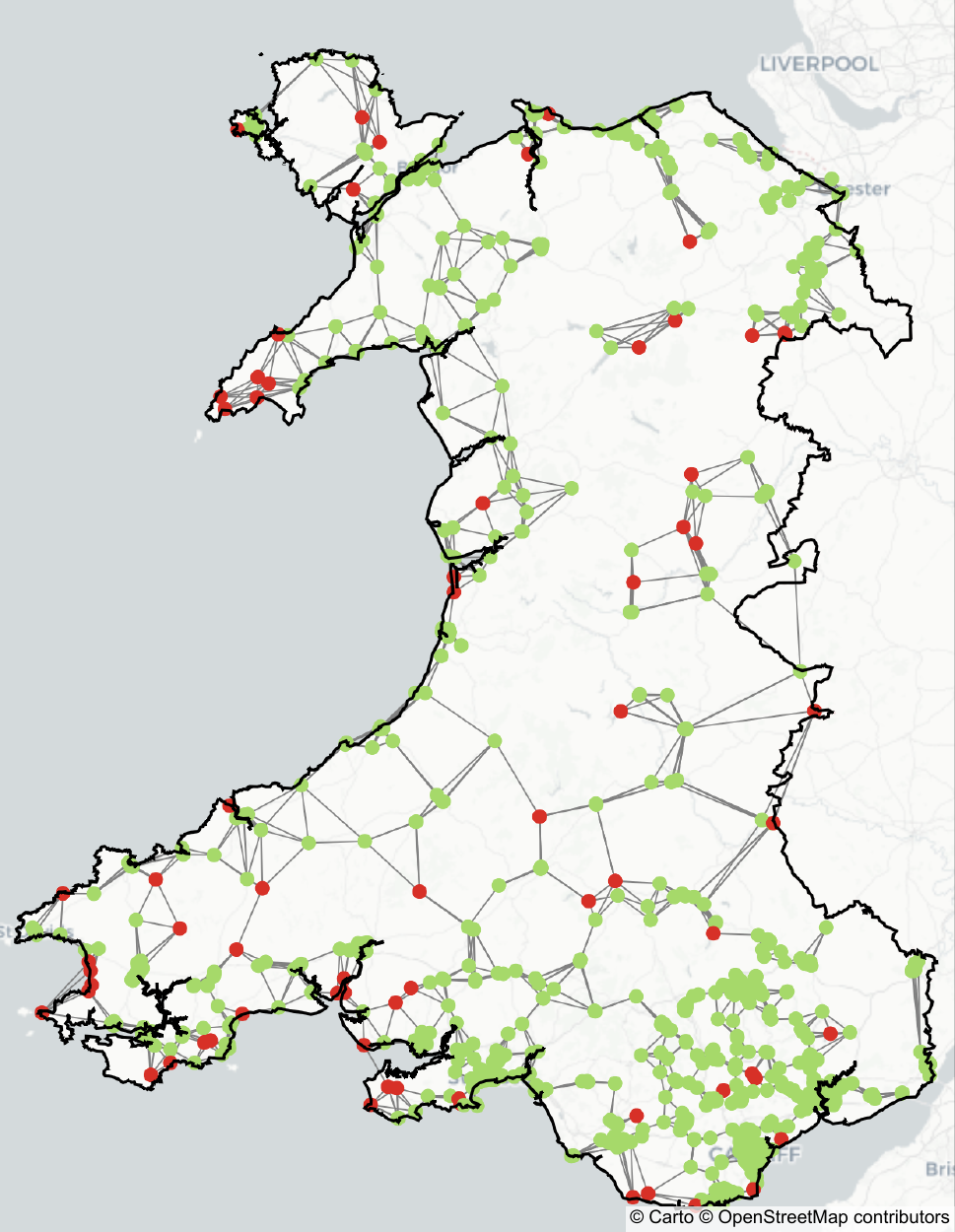}
        \subcaption{Road Proximity}
    \end{minipage}
    \hfill
    \begin{minipage}[b]{0.32\textwidth}
        \centering
        \includegraphics[width=\textwidth]{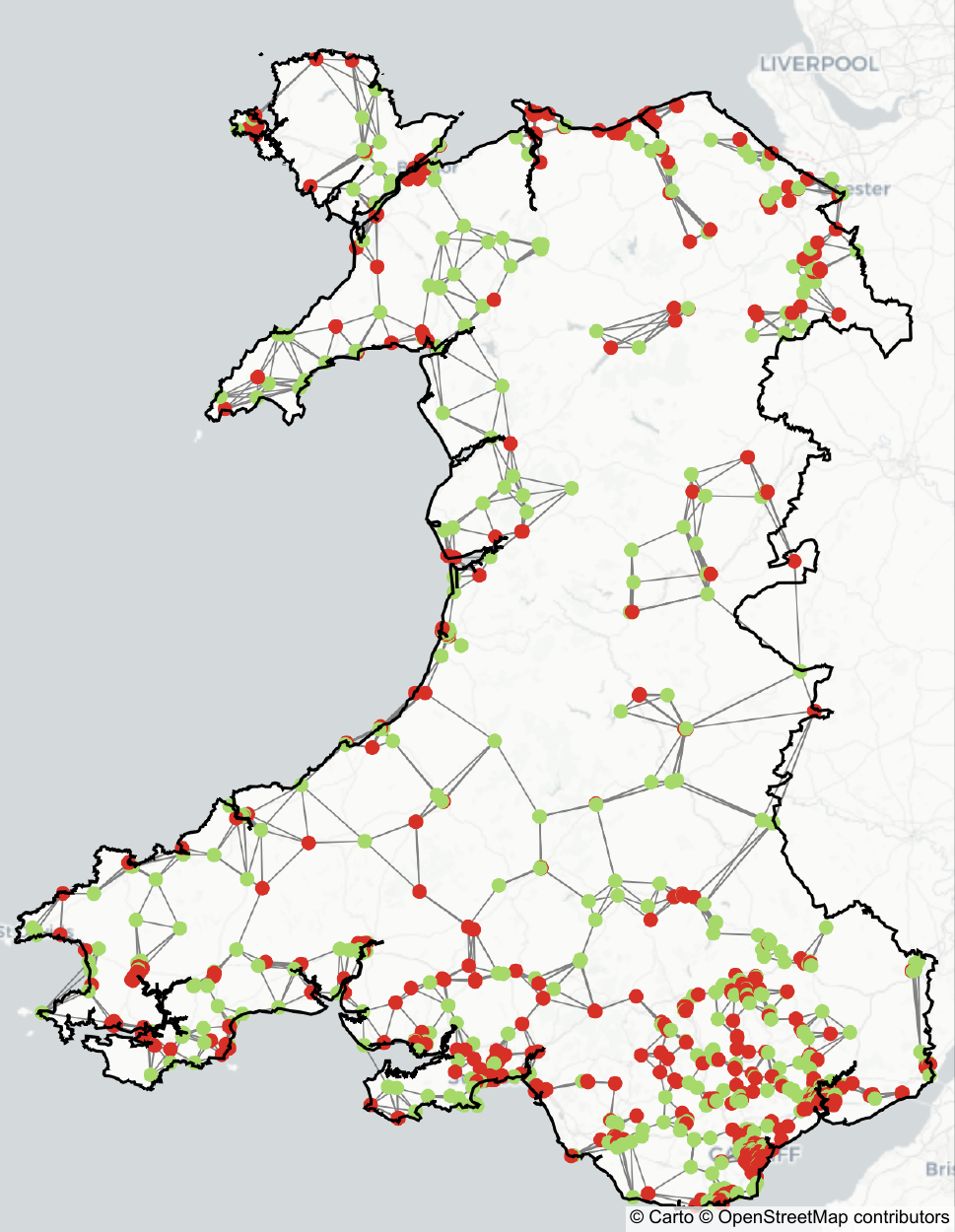}
        \subcaption{Vegetation}
    \end{minipage}
    
    \caption{Overview of EVCS and associated risk factors across Wales. Green markers indicate low/no risk, while red markers signify high risk. The graph view in (a) illustrates all EVCS as network nodes in blue.}
    \label{fig:evcs_risks}
\end{figure}

Figure~\ref{fig:evcs_risks} visualises the spatial distribution of individual risks across the charging network. The LST and vegetation risks show wide spatial coverage, with red-marked chargers clustered in southern and coastal areas. Flood risk exhibits more discrete lowland clustering. Road and grid risks appear highly localised, affecting only a few scattered chargers.

\begin{figure}[ht!]
    \centering
    \includegraphics[width=\linewidth]{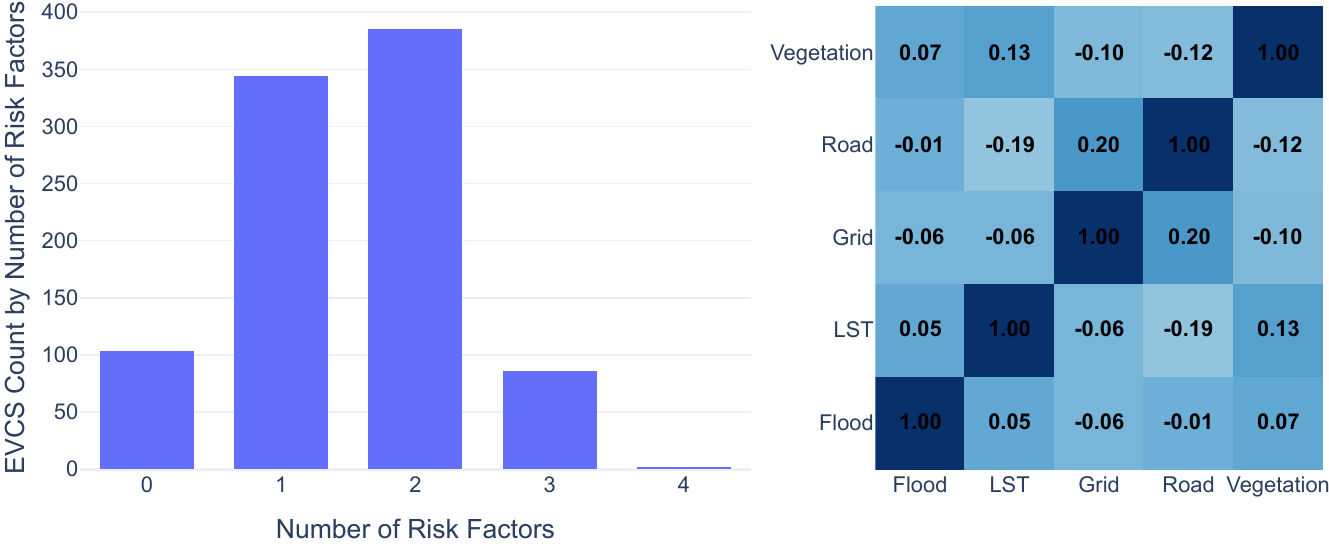}
    \caption{(Left) Distribution of EV charging stations by number of concurrent risk factors. (Right) Correlation matrix of binary risk indicators across all EVCS, annotated with Pearson coefficients.}
    \label{fig:factor_overlap}
\end{figure}

Figure~\ref{fig:factor_overlap} (left) shows the distribution of EVCS by number of overlapping risk factors. Most chargers fall into 2- or 3-risk zones, with a smaller share facing 4+ risk types. Figure~\ref{fig:factor_overlap} (right) provides a correlation matrix of binary risk indicators using Pearson coefficients. Weak to moderate positive correlations are observed between LST and vegetation, while other risk types appear largely independent.

\begin{figure}[ht]
    \centering
    \begin{minipage}[b]{0.53\textwidth}
        \centering
        \includegraphics[width=\textwidth]{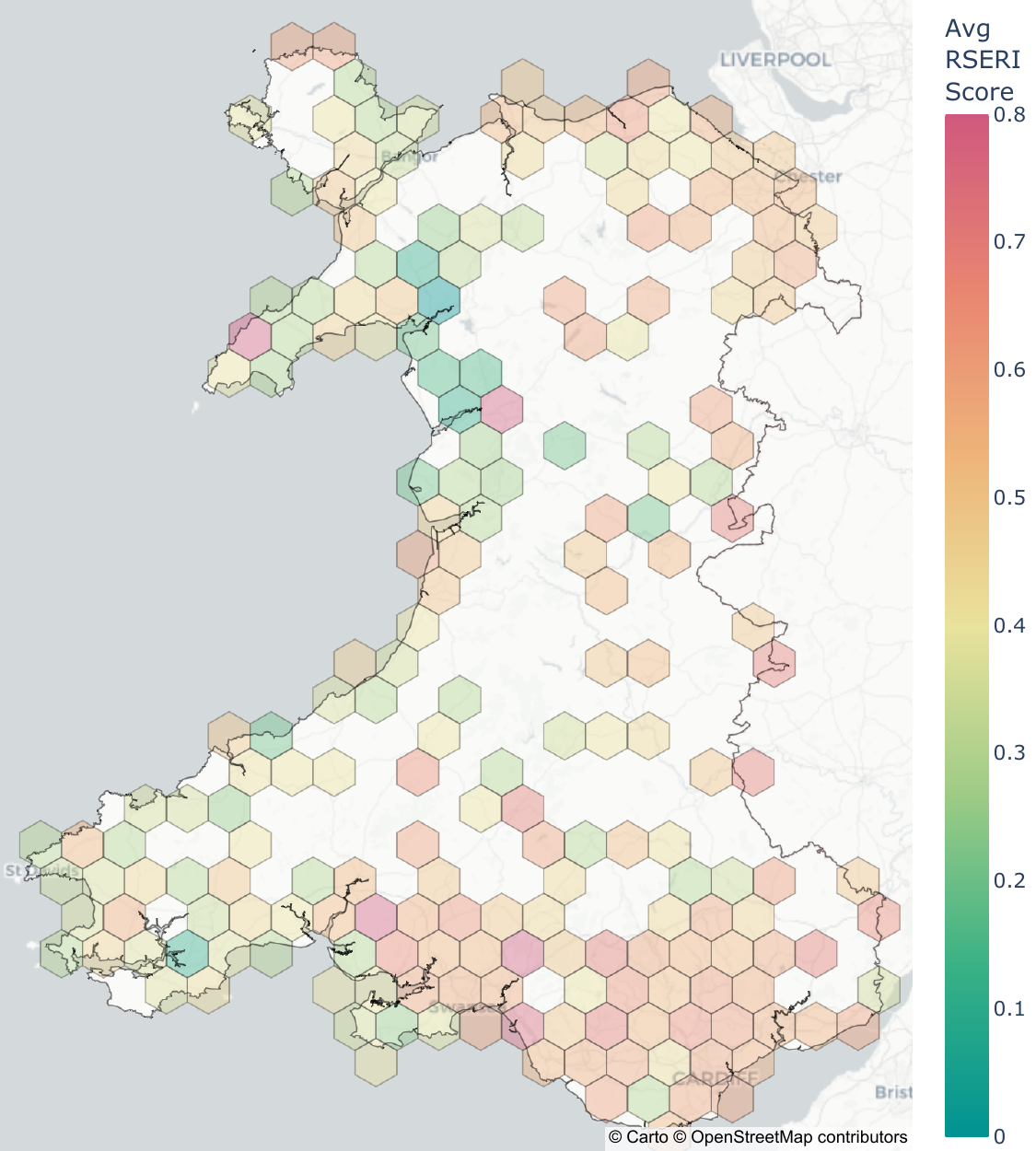}
        \subcaption{RSERI Score Distribution}
    \end{minipage} 
    \hfill
    \begin{minipage}[b]{0.46\textwidth}
        \centering
        \includegraphics[width=\textwidth]{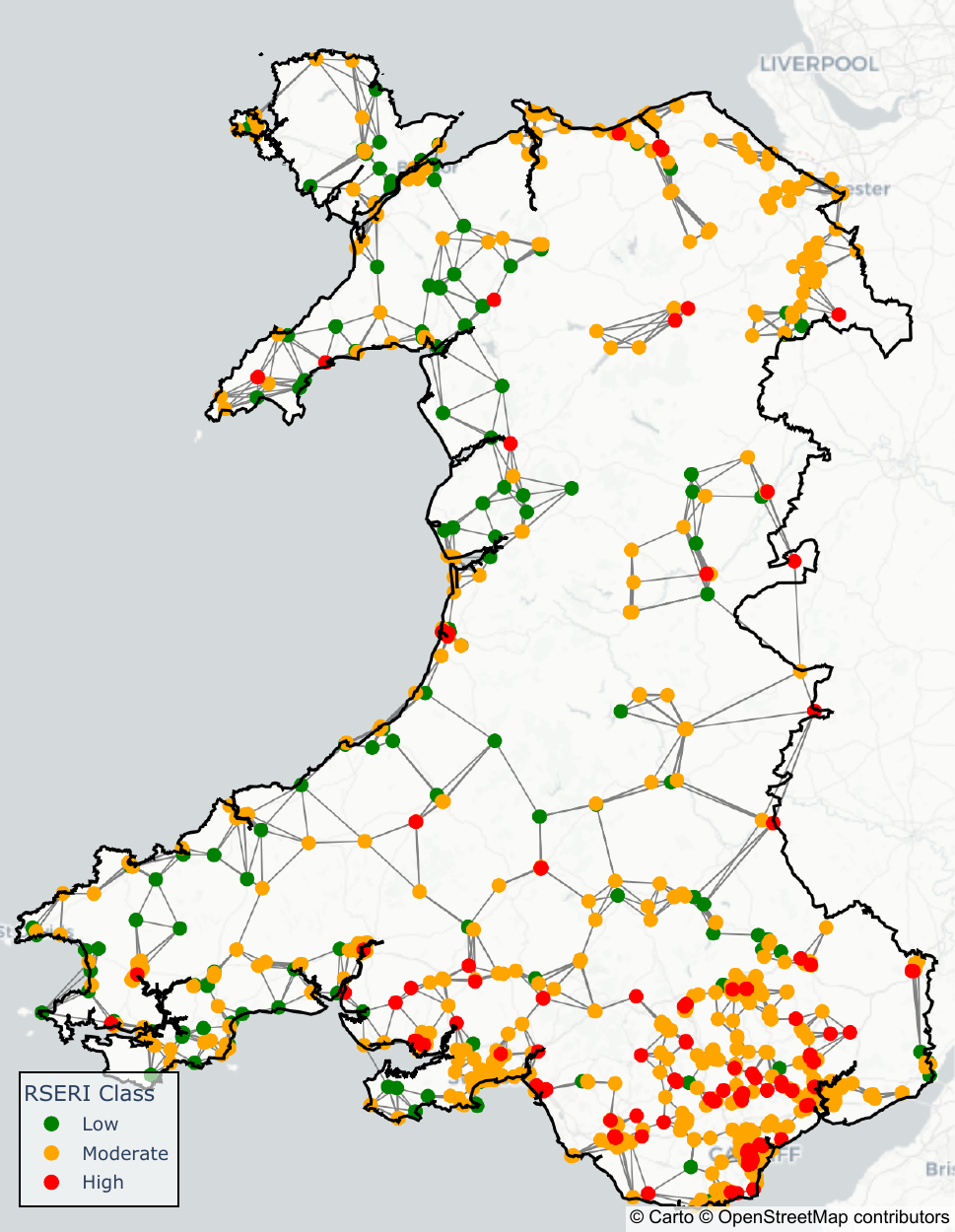}
        \subcaption{RSERI Classes Distribution}
    \end{minipage}    
    \caption{(a) Spatial distribution of average RSERI scores across Wales visualized using a hexagonal binning.
(b) Graph-based representation of EV charging stations colored by RSERI class (Low, Moderate, High).}\label{fig:rseri_distribution}
\end{figure}

Figure~\ref{fig:rseri_distribution} presents the spatial distribution of RSERI scores. The subplot (a) uses a hexbin visualisation to highlight the regional intensity of composite risk. Higher scores are concentrated in the South Wales corridor and major transport areas. The subplot (b) shows EVCS coloured by RSERI class (Low, Moderate, High) within a spatial graph. This highlights clusters of vulnerability and supports topological comparisons.

\begin{figure}[ht!]
    \centering
    \begin{minipage}[b]{0.56\textwidth}
        \centering
        \includegraphics[width=\textwidth]{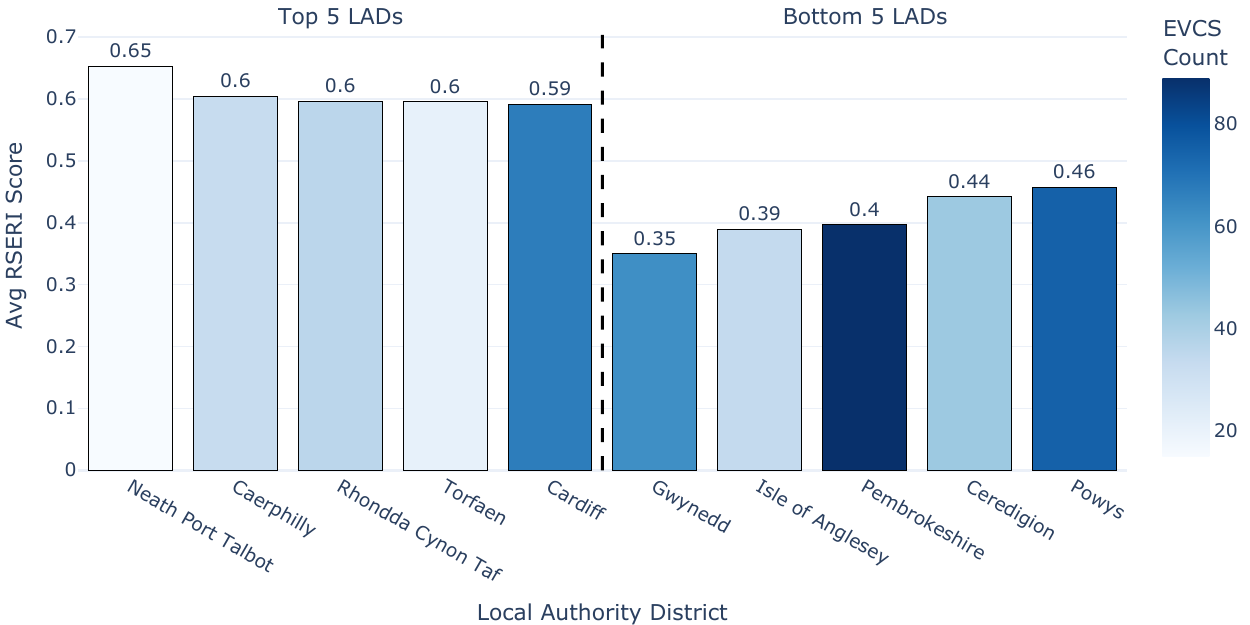}
        \subcaption{Top \& Bottom 5 LADs by RSERI}
    \end{minipage} 
    \hfill
    \begin{minipage}[b]{0.43\textwidth}
        \centering
        \includegraphics[width=\textwidth]{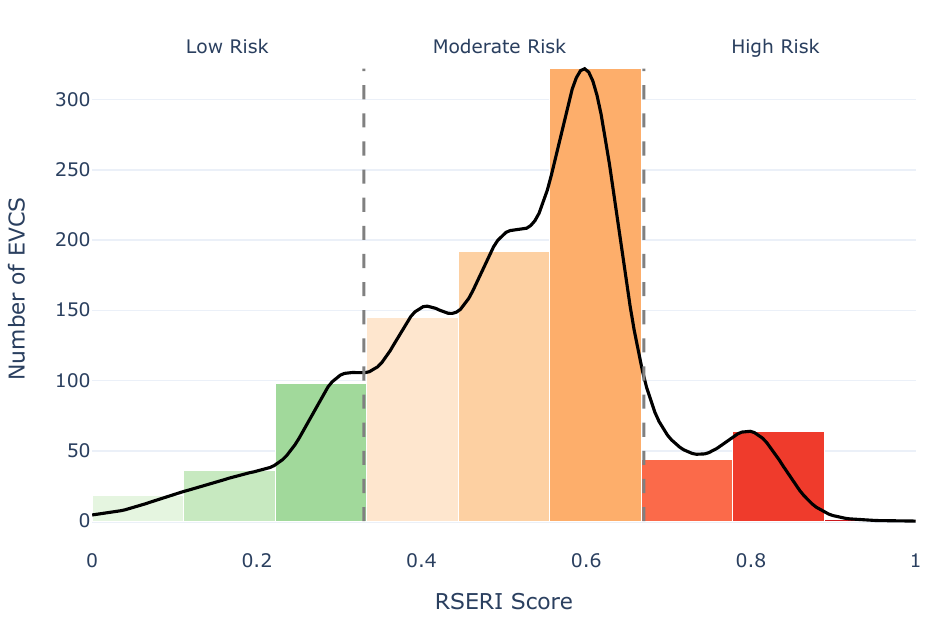}
        \subcaption{RSERI Score Histogram}
    \end{minipage}    
    \caption{
RSERI scores across Wales: (a) Top and bottom 5 LADs by average score, bar colour shows EVCS count; (b) Score distribution with KDE and risk class bins.
}\label{fig:lad_summary}
\end{figure}

Figure~\ref{fig:lad_summary} (a) ranks the top and bottom 5 Local Authority Districts by mean RSERI. Neath Port Talbot and Caerphilly show the highest average scores ($\geq$0.6), while Gwynedd and Pembrokeshire rank among the lowest. Figure~\ref{fig:lad_summary} (b) provides a histogram of RSERI scores, overlaid with a KDE curve and background class bins. The distribution shows a concentration in the moderate-risk zone, with a smaller high-risk tail.

\section{Discussions}\label{sec:discussion}

The RSERI-EV framework offers a reproducible and interpretable approach to quantifying climate and infrastructure stress at the spatial level of individual EV charging stations. By fusing diverse geospatial and remote sensing datasets, this initial implementation enables holistic resilience screening without requiring field-based or proprietary inputs.

The results show that the majority of EVCS in Wales are exposed to at least one form of environmental or infrastructural stress. LST- and NDVI-derived indicators capture widespread surface and thermal vulnerability, especially in urban centres and industrial corridors. Meanwhile, more infrastructure-focused risks such as grid and road proximity tend to be highly localised. These dimensions rarely co-occur, supporting the case for multi-modal modelling.

The RSERI score, as a composite index, effectively integrates these factors and provides actionable spatial insights. Hexbin visualisation of RSERI scores highlights risk clustering in South Wales and coastal regions, while the graph-based view offers a topology-aware perspective of network stress. Such representations could inform priority assessments for reinforcement or relocation under future grid load scenarios.

At the regional level, the analysis reveals substantial inter-LAD variability. For instance, Neath Port Talbot and Caerphilly exhibit consistently high composite scores driven by thermal and vegetation factors, while more rural districts such as Gwynedd present lower exposure. This divergence underlines the need for geographically tailored resilience strategies.

A limitation of this study lies in the binary treatment of risk indicators, which simplifies underlying gradients and uncertainty. In future work, continuous variables could be incorporated using fuzzy logic or probabilistic approaches. Additionally, while spatial graphs were constructed, their potential for downstream reasoning or optimisation was not yet fully explored. Network-aware metrics or graph-based simulations could enhance the decision-making in the future work. Finally, although this analysis was scoped to Wales, the methodology is generalisable. As more high-resolution EO and infrastructure data become available, the RSERI-EV framework could scale to other geographies or be used in dynamic simulations under climate scenarios.

\section{Conclusion}\label{sec:conclusion}

This paper introduced RSERI-EV, a spatial resilience screening framework for electric vehicle charging networks. By integrating remote sensing data, infrastructure attributes, and graph-based modelling, we quantified multi-hazard exposure across the Welsh EVCS landscape. Our analysis revealed concentrated vulnerabilities in urban and coastal districts, with heat and vegetation stress as dominant drivers. The proposed RSERI score offers a scalable, interpretable metric to support infrastructure planning under climate risk. Future work will explore dynamic risk modelling, continuous indicators, and integration with network simulation tools to enhance strategic deployment and adaptation.

\bibliographystyle{splncs04}
\bibliography{refs}
\end{document}